\documentclass[conference]{IEEEtran}
\IEEEoverridecommandlockouts
\usepackage{cite}
\usepackage{amsmath,amssymb,amsfonts}
\usepackage{algorithmic}
\usepackage{graphicx}
\usepackage{textcomp}
\usepackage{xcolor}
\def\BibTeX{{\rm B\kern-.05em{\sc i\kern-.025em b}\kern-.08em
    T\kern-.1667em\lower.7ex\hbox{E}\kern-.125emX}}
\begin{document}

\title{The Social Responsibility of Game AI\\
\thanks{Funded by a Royal Academy of Engineering Research Fellowship}
}

\author{\IEEEauthorblockN{Michael Cook}
\IEEEauthorblockA{\textit{School of Electronic Engineering and Computer Science} \\
Queen Mary University of London\\
mike@possibilityspace.org}
}

\maketitle

\begin{abstract}
Over the last decade we have watched as artificial intelligence has been transformed into one of the most important issues of our time, and games have grown into the biggest entertainment industry. As a result, game AI research as a field has enjoyed increased access to funding, exposure in the press, and influence with governments and some of the largest technology firms in the world. At this pivotal moment in the history of our field, this paper argues that this privileged position brings with it an important set of responsibilities which we have largely failed to meet. We show to whom we are responsible, identify some of these responsibilities, and suggest actions we can take as a community to leverage this power for good.
%
\end{abstract}


\section{Preface}
In this paper we discuss the social, political and cultural context in which game AI research is done today, and ask what responsibilities we have as researchers beyond our personal goals and our employers' desire for citations, press and money. We describe some of the major groups to which we have a responsibility today; we discuss the nature of some of these responsibilities; and we offer proposals for how we can meet these responsibilities in the future by changing what we do, how we do it, and who we work with and for.

Many of the arguments laid out in this paper are not new to us as a community. They are conversations we have on social media, at conferences after hours, and in our research groups. This paper is an attempt to push some of these discussions into the spotlight, to help preserve these issues in the permanent record of this field, and also to help illustrate the author's own views on these topics. We do not expect any reader to agree with everything; but we do expect everyone to take these conversations seriously. Many have been ignored for too long.



%


\subsection{A Note On Statistics}
Throughout this paper we reference the size of markets, billions of dollars invested or earned, or other financially-focused metrics. Given that an overemphasis on commercial gain is one of the issues we discuss, we wish to point out that we are not using such statistics as proxy for value or worth. However, in light of the emphasis placed on profit by governments and private companies, money is often a factor influencing their decision-making, and thus such statistics are used to indicate what may drive the actions of these organisations in the future.

\section{Do We Have A Responsibility?}
We begin by establishing where our unique responsibilities come from. In this section we introduce the two key subjects of this paper -- AI and games -- and illustrate their importance in everyday modern life, and where that places us as researchers.

\subsection{Games and Play}
By `game' we would include all digital videogames and boardgames, as well as adjacent and allied media like interactive 3D artworks and VR experiences. We also include folk games, playground games, and the simple act of undirected play itself that we engage in daily without thinking \cite{shonan}. 

%

Games and play are vital parts of our lives, and our appreciation of their importance in a healthy life is growing. Globally, the games market is estimated to be worth more than \$179.7bn \cite{idc}, with 2.7bn people buying and playing commercial games regularly \cite{gibiz}. This is a considerable underestimate of the importance of play in everyday life, as it focuses solely on commercial activity, but it underlines the significance of games not just in our daily lives but also in the global economy. The market for games in the UK accounts for more than half of all entertainment spending \cite{gama}, and globally the economic value of the games market is approaching twice that of film \cite{film}.

Play is a vital part of a child's development, helping them ``think, feel, and explore safely themes... such as justice, integrity, anger, jealousy, violence, compassion, difference, cultural variation and diversity'' \cite{bruce}. Accordingly, games are often thought of as being most important to younger generations; in a 2019 survey of UK children, for example, 93\% were found to play games regularly\cite{ripoff}. However, increasingly we acknowledge that games are an important part of the lives of people of all ages. A 2019 survey in the US found that 50\% of people over the age of 65 who play games had been playing for 10 years or less. In \cite{kelley} Kelley notes that play is just as valuable for adults as it is for children. They observe: ``play is a critical component of [adult] mental health, personal relationships, and fostering greater social connections'', and points to the surging popularity of so-called adult colouring books or the rise of play-based museum exhibits as evidence of our increasing embrace of play as a part of healthy adult life. Games are not simply a big number with a dollar sign in front of it: play is one of the most fundamental things we do.




\subsection{Artificial Intelligence}
By `artificial intelligence' we include anyone who studies, uses or develops AI techniques and theory, including planning, heuristic search, machine learning, computational creativity, human-computer interaction and more besides. This also includes people who study the use and impact of AI, such as psychologists, sociologists, ethicists and historians. This definition does not require affiliation with a university or any professional status -- developers, hobbyists, modders, tinkerers and players all help move the field of game AI forwards, and so should be aware of the discussion laid out in this paper. However, we place a special emphasis on those whose primary employment is in AI research at universities, private labs, startups, or large game developers. We have been afforded privilege, influence and resources to work in this area, and as such we carry the greatest responsibility to be aware of these issues and work to fix or improve them.

The last decade has seen an explosion of interest, investment and activity within AI research. According to the AI Index 2021 report, in the year 2020 publications about AI accounted for 3.8\% of \textit{all} peer-reviewed research publications worldwide \cite{aiindex}. Between 2015 and 2020, the number of artificial intelligence publications on arXiv increased by a factor of six. This surge in research is being driven by both public and private investment. Estimating exact figures is difficult because both government funding and private equity investment figures are hard to obtain. Analysis of public investment data on CrunchBase, a directory of public and private companies, shows a steady increase in funding acquired by AI startups \cite{crunchbase}. Between 2010 and 2020, AI startups listed on Crunchbase raised over \$73bn. This reflects only a small portion of total investments, however -- a Deutsche Bank Research report estimates that 2018/2019 saw \$65bn of private investment alone \cite{dbreport}. We can see matching trends in the actions of governments. US government spending on AI in 2020 rose to \$5.9bn, and is expected to increase to at least \$7bn in 2021. Estimates of investment by the Chinese government range between \$2 and \$8bn, and EU states have also earmarked funding for AI research such as France (\$1.8bn between 2018 and 2020) and Germany (\$1.2bn until 2023).

The explosive growth of AI has ramifications beyond the scale of investments. AI is now part of pop culture: late night shows in the US have segments where they play games with AI assistants or read out AI-generated text \cite{heyrobot}, and filters, deepfakes and AI assistants are part of daily life. AI is influencing how countries are run, with private AI labs obtaining privileged access to sensitive government branches including healthcare \cite{nhsdeepmind}, policing \cite{guardianpalantir} and immigration \cite{mijente}. Yet there is clear evidence that these changes are happening faster than many are able to regulate, respond to, or simply keep up with. Governmental organisations worldwide are seeking advice as they attempt to legislate on a wide range of technological impacts, from labour to creativity \cite{euclaw}; demand for AI expertise greatly outstrips supply \cite{stateofai}; and globally popular opinion is divided about whether to be afraid of AI or hopeful \cite{hopes}. Amidst this uncertainty and concern, the AI industry continues to move fast, and break things \cite{keyes}.


\subsection{Game AI Research}
We find ourselves today, as game AI researchers, at the meeting point of these two important areas. On the one hand, a huge entertainment industry built on top of one of the most fundamental human activities; and on the other, a once-in-a-century technological revolution that the world is desperate to understand, exploit and, in some cases, defend against.

Games offer a way to engage with complex ideas in an abstract and controllable environment. For researchers this makes them an ideal testbed, because we can tackle tasks in simulated environments, where real-world risks and consequences can be mitigated or reduced. This also makes games a healthy place for people to explore and play themselves \cite{jos}, although games are not entirely safe simply by virtue of being abstract \cite{aidungeon}. Nevertheless, this overlap of interests makes games an important meeting point for AI research and the general public, and is why private labs like Google DeepMind and OpenAI have targeted games -- they engage the public, they provide clear demonstrations for legislators and investors, and they satisfy the needs of AI researchers for tough challenges. 

In this light, we can see that games are not just a good benchmark for AI research, nor are they just a high-earning entertainment industry or a creative outlet -- they are an important political, social and strategic tool \cite{smithmartens}. While we might look at larger conferences, with tens of thousands of attendees and ballooning citations, working on `serious' applications of AI to medicine or language, and consider our work less significant, it is our belief that we are in a unique position to impact the world in a way that few working in AI can. Crucially, this leaves us with a set of responsibilities that no-one else can pick up, and means that the impact we have on the world can just as easily be negative as positive.


\section{To Whom Are We Responsible?}

\subsection{To Game Developers}
The games industry employs hundreds of thousands of people around the world -- over 220,000 jobs in the US alone rely on its games industry \cite{esa}. Yet we typically hear from just a fraction of those people, giving talks at major industry conferences, in high-profile interviews, or on stage at corporate press events. Our research touches far more people than this most visible few though, and we should recognise and understand better their needs, their perspective on our work, and their hopes for the future. In recent years several important issues have become more prominent in discussions about the wider games industry: stories of toxic work environments \cite{vergeubi}, widespread crunch and mismanagement \cite{peoplemakegames} \cite{polygon-crunch}, and growing support for unionisation \cite[p.~91-102]{woodcock}.

As game AI researchers, one of our most commonly-cited use cases or expected beneficiaries from our work are commercial game developers. This is reflected in the organisations our community engages with: in 2019, for example, COG's sponsors consisted of two game developers, one developer/publisher, and one middleware tools developer \cite{cog19}. Its Industry Day consisted solely of talks from commercial games companies, with three of its four keynotes coming from billion-dollar companies. Yet, despite our frequent claims about applicability to the games industry, and our efforts to build bridges to commercial developers, papers are often unclear on the expected ramifications of their impact on the actual workers in the industry, or use vague catchphrases and euphemisms about the potential benefits of our work which are, to our knowledge, completely unsubstantiated.

We have a responsibility to game developers because our work impacts their livelihoods. It has the potential to transform the kind of work they are asked to do, to transform the scale and type of projects they are assigned to develop, and in some cases may threaten to eliminate their job altogether. 

\subsection{To Artists, Hobbyists and Others Excluded}
In the year 2010 a total of 276 games were released on Steam, the most popular digital storefront for selling PC games. In 2020, 9,913 games were released \cite{steamspy}. This increase in the rate of games being released has put pressure on commercial developers as they struggle to stand out in a more competitive storefront \cite{polygon-indie} \cite{indiepocalypse}. However, this apparent crisis obscures a more important fact: the overwhelming majority of people making games today are not sold commercially, or do not have access to these `oversaturated' marketplaces at all. 

In February of 2021, Steam's storefront listed its 50,000th game \cite{steam50}. itch.io, a storefront with more open access requirements and no listing fee, hosts over 370,000 games as of April 2021, 357,000 of which are free \cite{itchgames}. Roblox, a game platform popular with younger players, has over 20,000,000 listings in its game store \cite{robloxgames}, all of which are free-to-play\footnote{Some of these games use in-game purchases, but we were unable to get data on what proportion. However, we estimate a vanishingly small percentage of Roblox games make money due to the mechanics of the store.}. The artists and hobbyists that make these games do so for fun and for self-expression, and their work is no less important for being noncommercial. Beyond this, there are many developers who have dreams of commercial viability but are unable to achieve this -- for example, because international sanctions block their access to global platforms such as Steam; because the capital required to enter the market is so high; or because they are not privileged enough to receive the exposure and networking opportunities afforded to others.

As game AI researchers, our impact on the games industry affects those not making money through development as much as it affects those who do. By researching techniques that can only be leveraged by companies with large budgets, we implicitly make it harder for smaller developers to compete. By not open-sourcing our work, or by releasing software incompatible with popular free tools, we make it harder and costlier to access and benefit from our work. Even the way in which we talk about outreach and engagement with `the industry' is loaded -- we tend to mean large, commercially viable companies, rather than the hundreds of thousands of people making games as part of their personal creative practice. We must talk to and understand the needs of these excluded and ignored developers, to understand and engage with them, in order to ensure we are supporting them, too.

We have a responsibility to these developers because our research affects the landscape in which they create -- the tools, techniques and standards that different people in the industry have access to. We stand outside the established power structures of the games industry, and are given significant amounts of funding, usually drawn from public money. As a result, we have an enormous opportunity to empower certain people, and to help those who are typically ignored, pushed away, or deprived of opportunities. If we truly believe in games as a powerful creative medium and an important form of personal expression, then we need to acknowledge our responsibility to everyone who participates in it -- not just the ones that contribute to economies in the Global North.

\subsection{To The General Public}
Artificial intelligence is rapidly transforming every aspect of our lives, and often not for the better. A 2019 survey of 150,000 people in 142 countries found that 30\% of respondents believed that artificial intelligence would `mostly harm' people, with some regions (such as Latin America and North America) reaching close to 50\% \cite{gallup}. Some more specific surveys put the figures much higher - a Pew survey focusing specifically on US views on job automation found 70\% of respondents fearing the impact of AI \cite{pew}. 

People encounter AI more and more frequently in their daily lives, yet often in situations where they are not in control, or where the stakes are high. AI is involved in, or becoming involved in, border control, policing, the justice system, hiring and firing processes, exam proctoring and student assessment, workplace monitoring, medical analysis and many more sensitive and important parts of our lives. Yet this rapid influx of AI into our lives has not been met with a better sense of understanding -- a 2019 survey conducted by the UK government found that only 1 in 10 people believed they knew a lot about AI, and more than half felt they did not know the impact AI would have on their lives \cite{ukgovunderstanding}.

Games are one of the few spaces in which people can encounter and learn about AI in a controlled way. We can already see how desperate people are for this kind of understanding through the impact of AI that provide similar affordances \cite{heyrobot}. Throughout popular culture we see a specific, corporate-focused image of AI being depicted, which narrows the public's ability to imagine different futures \cite{oii}. A better public understanding of AI is required to help the public determine the kind of future they want, or what changes they wish to resist. Games are an important medium for promoting this understanding, and one which we can have a massive impact on, by opening up our research, promoting accessible ways to learn about common AI techniques and systems, and showing people what other things AI can do.



We also influence and shape the technology powering one of the most popular leisure activities in the world. While we think of game AI primarily as a way to make games more fun or interesting, it also has more serious ramifications for players -- our research affects how games track and model player behaviour and engagement, for example, and how this subsequently can be monetised. The impact of loot boxes and microtransactions has become a point of public concern over the last decade \cite{gambling}. And beyond games, our work fuels developments that affect the public in serious ways, including the development of technology with military applications, as we discuss later.

We have a responsibility to the general public because our work will shape the future of a major source of entertainment and creativity for them, and because game AI is a vector for many other kinds of AI research that affects other parts of their lives. But on top of this, we have a responsibility as scientists working in a uniquely engaging and accessible space, to help equip the public with the knowledge they need to navigate and survive this new wave of technology.

\subsection{To Each Other}
According to a list compiled by Mark Nelson, based on the proceedings of fifteen major technical games research venues, there were over 2,140 people active in the technical games research community between 2011 and 2021 \cite{nelson}. 495 of these are considered `regular' community members, publishing two or more papers or thereabouts\footnote{Mark uses a more nuanced definition involving fractional authorship for multiple-author papers; a regular member has 2.0 papers or above.}. In addition to our responsibilities to the many lives our research impacts in the wider world, we must also consider the responsibilities we have to those in our community, those who have been forced to leave, and those who are yet to join.


The problems faced by researchers in academia are numerous and well-known. Racism and sexism are rife within the academy \cite{racismuk} \cite{rs-race} \cite{basford}, for example, problems which are compounded by the precarity of the career ladder, making it difficult to pursue a career without moving for jobs, taking pay cuts, delaying major personal life decisions and often damaging physical and mental health through overwork \cite{naturedepression}\cite{burnout}. Many of these issues are particularly acute for game AI -- our research is often at the fringes of fields, making it harder to find jobs, get promoted, or secure funding, and issues such as racism and sexism within games amplify the existing problems in academic communities. Yet none of these issues are insurmountable. Some require facing down powerful institutions: universities, publishers, funding agencies or governments. Many more of these issues we perpetuate against ourselves, through our support of these broken systems, or in our inaction to fix the problems we see around us.

We have a responsibility to each other because we are a community -- a community not defined by rankings, earnings or influence, but by our shared interests and beliefs. Our community is spread out and constantly shifting, and includes those in private labs, universities, games companies, tech firms, and on their own as hobbyists and developers. We are all responsible for looking out for one another, and for leaving this field in a better state than we found it.

\section{What Responsibilities Do We Have?}
In this section we identify some of the issues we are facing today as a community. In each case, we outline the issue and then propose some actions that could be taken today.

\subsection{Resisting Imperialism and War}
The link between games and the military is long-running. Games such as \textit{America's Army}, developed by the US Army itself, are used both for recruitment and training, alongside games such as \textit{Operation Flashpoint} and its successors in the \textit{ARMA} series \cite{arma}. In a push for realism, game developers liaise directly with arms manufacturers and in the past have even provided links on official game sites to purchase guns used in their game \cite{armsdealers}. In recent years we have seen an even stronger push to use games for PR purposes, such as the US Army's (largely catastrophic) attempts to start an eSports team \cite{armyesports} and stream on Twitch \cite{armytwitch}.

In 2021, the United States' National Security Commission on Artificial Intelligence released their final report, a 756-page document of recommendations for `winning the artificial intelligence era' \cite{schmidt}. The first example of an AI breakthrough used in the report is Google DeepMind's AlphaGo system. Governments around the world believe that AI is vital to the future of their military and strategic goals, and also understand the important role that games play in this sector. Games are already used as training for people, and soon they will be used as testing grounds for more complex systems, including autonomous weapons. Robert Work, one of the NSCAI's report coauthors, has described the development of autonomous weapons as a `moral imperative' \cite{imperative}.

The link between the military and game AI research similarly stretches back into our history, with the most common example being the use of DARPA funding for our US-based colleagues, with varying levels of connectedness to military applications. In a survey of five years of papers both at IEEE CoG and AAAI AIIDE\footnote{We originally intended to survey the Foundations of Digital Games conference but they are still paywalled, a fact which is incomprehensible in itself in the year 2021.}, we found 28 publications either funded by military organisations or with co-authors bearing military affiliations. Of these, 15 were in the last year. This is likely to be an underestimate as the majority of papers do not state their funding, links to militaries are often obscured, and in some cases connections are intentionally kept secret \cite{guardianmi5}.

We find the presence of military organisations and funding bodies at a games conference grossly inappropriate at best, and at worst, complicity with killing and oppression. For military organisations, papers and conference talks have a similar impact to games, in normalising the presence of the military in our lives, supporting recruitment (in this case, of students and researchers) and promoting a sanitised image. Furthermore, they ask our community members to spend their time providing feedback and critique to further the aims of military organisations, supporting the very systems of imperialism and war that harm not only the groups to which we are responsible, but our own research community. Many of our colleagues are unable to attend conferences, publish papers or otherwise conduct research because of the foreign policies of countries like the United States; while others have had their lives touched by the impact of military action around the world. Some members of our community have been forced to flee the actions of the same militaries whose funding and research we welcome with open arms -- an insulting state of affairs. This is not simply a question of scholarly community-building, but a question of whether we want our legacy as a field to include supporting the military-industrial complex.

Government funding of artificial intelligence, especially in Europe and America, is likely to continue to increase over the coming decade. At the time of writing, the Future of Life institute has identified 36 national AI strategies and six international strategic alliances including the EU and the UN. Many of these strategies involve escalating levels of investment, and increasingly the spending on AI R\&D is directed towards military ends. In 2020, for example, over 80\% of US Government spending on AI was on defence. In the UK, the Defence Science and Technology Laboratory has expanded its AI remit considerably in recent years, and acts as a partner in current games research grants. We are sleepwalking into a dangerous situation in which we enable and support a new era of warfare and imperialism, while simultaneously helping launder the reputations of those responsible for it.

\textbf{Actions}
In order to push back against the encroachment of military and defence interests in our field, we believe that immediate action should be taken to limit the presence of military-oriented research at our venues. In particular:

\noindent \textbf{1)} Do not accept conference or journal submissions from authors working for, or affiliated with, intelligence agencies, military schools, branches of military organisations, or other defence-related institutions and companies.

\noindent \textbf{2)} Do not accept conference or journal submissions describing work whose stated real-world applications or aims are militaristic in nature -- this includes AI for strategic planning or command, and autonomous weapons.

In addition to this, we recommend encouraging open discussions about the presence of military connections in our work, and to help us move away from this as a community.

\noindent \textbf{3)} Require in publications disclosures of military links through funding that is not explicitly from military sources (e.g. a government-backed grant where a defence institution is a partner or where applications include specific military outcomes).

\noindent \textbf{4)} Support one another to help transition away from defence-based funding sources, even where the research is not explicitly military in nature (e.g. `non-military' DARPA grants).

We are aware that this issue, as with everything else in this paper, is entangled with other issues. Grant money keeps people employed, which is important for a host of reasons, including the maintenance of work visas. We are not suggesting with item 4) that our reliance on funding from agencies such as DARPA can be resolved overnight. Some of these actions represent ideal end points, and will take time to work towards.

\subsection{Resisting Capitalism}
The impact of capitalism can be felt everywhere today, in every major news story, from the brutal impact of vaccine patents, to the rapid acceleration of climate change. As game AI researchers we influence, and are influenced by, capitalist systems in all of our work. The negative impacts of this on those we are responsible to are too numerous to list here, and so in this section we chose to highlight two specific examples.

\subsubsection{The Tyranny of Scale}
The more capital possessed by an individual or organisation, the proportionally higher returns they are able to acquire on their investments. This vicious cycle leads to the accumulation and concentration of capital in a smaller and smaller group over time, simultaneously forcing out competitors and entrenching their control over those with no ownership of capital at all. Over the last decade, we have witnessed this effect greatly impact AI research, as private capital took an interest in large-scale machine learning.

In 2019 OpenAI published an assessment of advances made in AI research \cite{openaicompute}. They found that, prior to 2012, the computing power required for a major AI result approximately followed Moore's law, in that it doubled every two years. Between 2012 and 2018, however, it doubled on average every 3.4 months. The computing power used for a major result increased by a factor of 300,000 between 2012 and 2018. Estimates for the training cost of recent systems include GPT-3, which is estimated to have cost \$12m, and AlphaGo Zero, which cost around \$25m. This is without considering the huge environmental cost of large-scale compute \cite{climate}.


The enormous capital involved in such landmarks makes working in the same areas as large corporations highly volatile -- something that game AI researchers working on Starcraft 2, DOTA or Go have experienced first-hand. Competition in these cases is not a matter of ingenuity or invention, but a matter of investment. Capital itself is the key innovation offered by such research, the secret ingredient that provides cutting-edge progress. As a result, major technology firms are engaging in research that only those with capital can profit from, repeating the patterns of concentrating and accumulating capital that we see replicated elsewhere in our economy and society.

Advocates for this research point out that cutting edge technology is often inaccessible, but becomes more efficient over time. Another OpenAI report published in 2020 supports this, pointing out that the cost of training a neural network on a specific image classification task has decreased by a factor of 2 every 16 months \cite{openaiefficiency}. Yet this is still less than the scaling shown in the 2019 report, and only focuses on the most common neural networks task. It also fails to solve the issue of stratification -- the richest will always have access to and control over the best, while the remainder of the world waits for advances to trickle down to them. This is also why another common defence -- that trained models can be efficiently distributed via cloud services as seen with GPT-3 -- also do not hold water, because they centralise control of important AI technology within the few corporations rich enough to create them. We need look no further than OpenAI, a supposed non-profit that sold exclusive access to their million-dollar language model to Microsoft at the first opportunity.

To reject the tyranny of scale in AI we need to do more than simply resist large-scale machine learning: we must actively work to develop research that is accessible, scalable, and cheap. Universities have responded to scale by trying to compete -- obtaining their own GPU clusters to do larger-scale work on, for example. Yet even researchers without such access must observe that the `ordinary' resources available to many of us reflect privileged access to technology available to a fraction of the world. The OECD reports that in 2017 97.6\% of households in the Netherlands had access to a home computer. As of 2019, the figure for Mexico is 44.3\%, and for Brazil it is 39.4\% \cite{oecd}. In 2016 the World Bank reported that access to the Internet was similarly unequally distributed \cite{worldbank}. Access in rich countries in the Global North is high, such as the United Kingdom where 94.78\% of people use the internet. Access in other countries is much lower - in Afghanistan it was 10.6\%, and in Somalia just 1.88\% of the population had used the Internet in the three months preceding the survey. Scaling up is not the correct response to capital's disruption of AI research -- we cannot solve the world's problems with cloud computing and high-performance clusters.

We must strive, as a community, to work for everyone. This does not mean we have to limit all research we do to within the least resources possible, but it does mean that we should incentivise research that works on smaller scales. In the face of a changing climate and an escalating crisis of consumption and resource scarcity, we should not do our research under the assumption that technology will always get faster and energy more plentiful. We should strive to make the research we do as accessible as possible -- both in how we distribute it, and in the resources and knowledge required to engage with it.

\subsubsection{Labour and Automation}
Capitalism relies on exploitable labour to survive and grow, and the exploitation of labour within the games industry is widespread. People working in the games industry around the world are subject to many forms of exploitation, including wage theft and crunch, and recently this has disproportionately been felt by workers in the Global South\cite{peoplemakegames}. A commonly-shared myth is that AI will make our lives easier by automating drudgery and mundane work. While this might be the case for those who are self-employed game developers, for many working in the industry there is no reason to think this will be the case.

Indeed, while it is common for game AI researchers to motivate their work by describing how it will make game development `easier', `faster' or `cheaper', the answer to one question is often missing: \textit{for whom?} For example, procedural generation is often cited as a way to make game development easier by automating content creation \cite{smithpcg}. However, there are several notable examples of commercial tools which leverage generative techniques and have achieved widespread adoption within the games industry, such as Substance Designer \cite{substancedesigner}, Houdini \cite{houdini} and SpeedTree \cite{speedtree}. Yet there is little sense that crunch or worker exploitation has reduced as these tools have become more widespread. What we do know is that games have gotten more complex and expensive, teams have ballooned in size, and the pressure put on them has increased.



Companies investing in new technology do not do so to reduce the workload on their employees, but to increase in the amount of work an employee can do in the same amount of hours \cite[p.~219]{marx}. Better tools and technology simply serve to change the \textit{nature} of work, and generally do not change the amount or intensity of it, unless making a job obsolete. In many cases we are willingly enabling the worst excesses of this, by engaging in research which aims to accelerate the pace of game development or eliminate certain jobs entirely. Technology is put to use in existing power structures. This means our research may, and perhaps inevitably will, be used to accelerate the extraction of value from workers and customers. We must do research with this in mind.

It is no longer acceptable for us to vaguely wonder about the impact of our research. We must take seriously the idea that game AI research impacts the lives of hundreds of thousands of workers around the world, and that if we choose to align our goals with capitalists, we are likely to end up harming more people than we help. This is not to say that we must stop doing any research that might be usable by large businesses. But we should at least be realistic about the impact of what we do, and discuss it with the same care and detail that we afford to descriptions of algorithms and experimental methodologies. In the long-term, I believe we should reflect on the nature of AI research and how it enables capitalistic processes, and to develop a theory of anticapitalist AI research that explicitly rejects and resists these in the nature of the work itself.



\textbf{Actions}
We cannot dismantle all of capitalism by doing our research a little differently. But we can stunt its advance in the industries we affect, and we can shift from actively supporting it, to helping others resist and survive it. 

\noindent \textbf{1)} Require disclosure of hardware and compute costs involved in obtaining experimental results and running systems.

\noindent \textbf{2)} Require impact statements with paper submissions, following the example of conference such as NeurIPS, with specific attention paid to the impact on automation of labour, acquisition and use of data, and player tracking and modeling.

\noindent \textbf{3)} Create specialised submission tracks for research which operates within technical resource constraints (e.g. fantasy platforms such as PICO-8) to encourage low-tech innovation, and tracks for optimising existing techniques and systems.

\noindent \textbf{4)} Support and expand existing artefact evaluation efforts, with incentives for open-sourcing work, and expand initiatives like the CoG Short Video competition that encourage outreach and communication of research.

In addition to these smaller steps, we also believe that there can be anticapitalist AI research. This is too large a discussion for this paper, but resisting the automation, surveillance and exploitation that our research enables can become a goal of our field, if we work to make it one.

\subsection{Building A Better Academy}
Academia's problems are numerous and well-documented, and we have already listed and cited some of them in this paper. Some of these problems are widespread across society, and ways to fix them are not unknown -- all that remains is the hard work of doing so. For example, while we cannot claim to be able to solve problems such as racism and sexism easily, we know that educating our communities, laying our clear rules for conduct and behaviour, and robustly enforcing those rules, help to produce safer and better spaces.

Academia also possesses many problems which are unique or at least less common. Many of these stem from two sources: an overly conservative adherence to tradition, and a deeply interconnected set of exploitative systems that govern our careers. Peer review, for example, is often held up as a tenet of the modern scientific process, and having articles peer reviewed affects hiring, promotions, grant funding and a host of other academic processes. Yet we also know that peer review is deeply flawed -- blinding is inconsistently used across conferences, the quality of reviews is highly variable, and a 2014 study of NeurIPS reviews showed that 57\% of accepted papers would have been rejected if the review process was rerun \cite{neurips}. Changing the process of peer review, or abolishing it in its current form entirely, is unthinkable for most academic communities.

However, our field has some advantages over other areas of Computer Science. Game AI research is young. If we limit our focus to the study of digital games in particular, most of the people who have ever done research into game AI are still alive today, and working actively. This makes us a young field, and one which sits outside mainstream research. Despite a burgeoning interest from the media and private research labs, research into `games' is often hidden under other terms in order to obtain funding, departmental support, or publication at larger conferences. 

We can use the outsider status and youth of this field to our advantage, however. Being less beholden to tradition allows us to change, to experiment and try new things that other fields would struggle to implement or convince its entrenched old guard to do. Being a less mainstream academic field allows us to take risks that might affect our metrics or prestige, because we already benefit little from such systems. We can experiment with new ways of evaluating and sharing knowledge, that deprioritise low acceptance rates and instead embrace community sharing and feedback. 

The actions outlined below in this section are the smallest, most minimal changes we could make, but we would urge particularly bold thinking here. We have let down far too many members of our community over the years: by not showing solidarity with them when they were hurt by those both in and outside of our community; by maintaining exclusionary policies and gatekeeping that has blocked access to the resources and networking required for academic careers; by allowing nepotism and cowardice to lead to inaction over abuse, exploitation and harm. We have lost so many brilliant and kind people due to our inaction, and we must work harder and be braver in changing our community -- not only for us, but also to provide an example for other academic communities that better systems and spaces can be built.


\textbf{Actions}
\noindent \textbf{1)} Collectively boycott conferences that do not publish a robust and enforceable Code of Conduct, or which fail to enforce it.

\noindent \textbf{2)} Collectively boycott conferences that are not open access, and that do not provide free public streams or recordings of talks (where authors agree). At the time of writing this would cover FDG, supported by the ACM, whose papers and talks are paywalled.

\noindent \textbf{3)} Collectively boycott conferences that do not allow remote presentation. Remote presentation should be allowed at any conference, without providing a reason, as long as relevant conference fees are paid.

Following jingoistic restrictions by the US in 2017, some US-based conferences allowed remote participation under certain circumstances. This became (temporarily) standard during the pandemic. Requiring explanations for remote participation is unnecessary and demeaning -- colleagues with family and care commitments, funding and visa issues, health and accessibility needs or any other range of cases should be supported. Remote access and participation must continue once this pandemic ends.

\noindent \textbf{4)} Create opportunities for academics to redistribute funding, for example by providing higher tiers of conference ticket packages that subsidise free, reduced-rate and student tickets. This allows academics who can justify higher spending to support the community if they feel able.

\noindent \textbf{5)} Create space for previously-published papers to be presented again (and possibly included in the proceedings) of conferences such as AIIDE and CoG, where the paper was originally published in a regional or national event in parts of the world that are underrepresented in our community. This will help raise the profile of scholars who have historically been unable to publish at or attend our conferences and build links to new research communities.

By calling for a boycott in some of these recommendations, we are not suggesting that this must necessarily come into effect by the 2022 conference season, as some of these changes may not be possible at this stage in the organisational process. But we must draw the line somewhere, and demand change.

\section{Conclusions}
We are currently experiencing a historic boom for AI research, that has afforded many of us increased funding, opportunities and impact. This hugely privileged position gives us an opportunity to effect real change -- within our own communities, within the games industry, and in the wider world beyond. Doing so will not be easy, and for many of us will mean risking funding, promotions, fame and opportunities, as well as burning bridges with friends, colleagues, industry connections, and employers. The best time to act, as they say, was yesterday. The second best time to act is now.

This paper is not an imposition of doctrine. It is simply, in the words of Keyes et al, a call for this community to either `justify the way things are, or join us in changing them' \cite{keyes}. To continue as we are is to endorse the status quo, to ignore the suffering of those around us, and to ignore our contribution to it. This paper is also not an exhaustive list of the problems experienced by our field, nor a comprehensive set of steps to fix them. It is the merest of starting points, with suggested actions that are, by and large, cheap and simple to implement. But they require us to act as one, as a community of equals, in solidarity with one another. The problems outlined in this paper did not appear out of thin air. They were built, little by little, by people like you and me. Similarly, the solutions will not appear by magic. They must also be built, little by little, and they can only be built together.

\section{Acknowledgements}
This paper is the result of countless conversations had between good friends on dark days, and I thank every single person who cares about these issues, who has stood up for them, and who will continue to work to fix them. Thanks to Jamie Woodcock, Os Keyes and Max Kreminski for specific feedback on the paper, and also thanks to Anndra Dunn, Sam Geen, Matthew Guzdial, Ian Horswill, Darius Kazemi, Chris Martens, Antonios Liapis, Amanda Phillips, Emily Short, Gillian Smith and Anne Sullivan -- a small fraction of the many people to whom I owe a great deal. Thanks to Alexia Revueltas Roux for providing invaluable in-depth citations about play and wellbeing. Thanks to Juan Mateos Garcia, Holly Nielsen and Youn\`{e}s Rabii for their input on the paper's direction too.

\bibliographystyle{plain}
\bibliography{biblio}

\end{document}